\documentclass[10pt,a4paper,showpacs]{article}
\usepackage[utf8]{inputenc}
\usepackage{amsmath}
\usepackage{amsfonts}
\usepackage{amssymb}
\usepackage{graphicx}
\usepackage[nottoc]{tocbibind}
\usepackage{authblk}
\begin{document}
\title{Transmitting a signal in negative time}
\author[1]{P. Singha Deo \thanks{deo@bose.res.in}}
\author[2]{Urbashi Satpathi}
\affil[1]{S.N. Bose National Centre, JD Block, Sec - 3, Salt Lake, Kolkata, India 700098}
\affil[2]{Raman Research Institute, C. V. Raman Avenue, 5th Cross Road, Sadashivanagar, Bengaluru, Karnataka 560080}
\date{\today}
\maketitle
\begin{abstract}
There is no self adjoint time operator defined in quantum mechanics. However, time intervals can be defined in several ways and can also be probed
experimentally. Our interest in this work is traversal time and signal propagation time. According to Copenhagen interpretation of quantum mechanics the
two should be the same but the issue is not settled yet in regimes where they can be negative. 
We use Argand diagram and Burgers circuit to show that the correct traversal time
and the correct signal propagation time can be identically negative implying signal can be propagated in negative time.
Some other physical consequences are discussed.
\end{abstract}


Time in quantum mechanics appear as a parameter and there is no self adjoint time operator consistently defined yet. This is not a serious problem as experimentalists
only measure time intervals and quantum mechanics give this measured time intervals very correctly in the semi classical limit \cite{land}. The low energy quantum
limit is confusing in 1D, 2D and 3D since calculated time intervals is not always
consistent with the Copenhagen interpretation of quantum mechanics itself \cite{land}. It is known that quantum mechanics
starts from a very different set of axioms and does not have to respect theory of relativity within the single particle coherence length \cite{land}. 
Within this coherence length it is well known experimentally and theoretically that a point particle can simultaneously be present at more than one spatial
positions and so a traversal time from one point to another point can be negative. But whether it can be negative in a well defined and physically consistent manner
is the issue still unknown and contribution of our work is to prove the consistency of negative times \cite{land}.
For this we have to go beyond 1D, 2D and 3D and look into quasi 1D (Q1D).
So when we say that negative times are possible then we do not make any reference to theory of relativity. But even within the axioms of quantum mechanics
the meaning of negative time has not been completely explained and we want to point out some examples to show that negative times are
completely consistent with quantum mechanics. 
There are two traversal times defined in quantum mechanics that are called Larmour precession
time (LPT) and Wigner delay time (WDT) \cite{land}. There are many other ways of defining time that correspond to different physical situations, 
but they cannot be called traversal time because they
are not consistent with the number of intermediate states through which a particle is propagated or traversed in time \cite{land}. The definition of traversal time
is that it has to be consistent with the density of states (DOS) upto a constant factor $\frac{\pi}{\hbar}$.
Different candidates for traversal time may have different mathematical forms but should lead to the same quantitative value inorder to be consistent.

In this paragraph we restrict our discussion to 1D.
The derivation of LPT \cite{butt1, butt2}
starts from a monochromatic plane wave or a stationary state and derives an expression for traversal time. Hence it starts from the time
independent Sc. eqn and derives a time. 
LPT is physically the average traversal time of electrons in the stationary beam and
such electrons in a stationary beam cannot be used to send a signal. If not restricted by having to carry a signal or information, even in classical physics
one can exceed the speed of light. A signal can be sent by using a wavepacket for example a large or a small wavepacket can mean different things.
Derivation of WDT \cite{land} actually considers the propagation of a wavepacket in time to derive a traversal time. The LPT
is exact but the WDT is approximate using stationary phase approximation.
If stationary phase approximation is valid at some energy range then the LPT and the WDT
are quantitatively the same, positive definite and fully consistent with DOS. 
This happens at high energies when quantum effects are moderate 
confirming the Copenhagen interpretation as will be explained below. 
At low energies when quantum effects dominate, then LPT and WDT do not give the same result. LPT is exact (as the derivation is exact
but explicit
calculations at low energies to show its negativity has not been done) while WDT is approximate and
can become negative (at these low energies WDT is no longer consistent with DOS and so its negativity need not mean anything physical).

We describe below the derivation of WDT as we want to clearly outline the nature of the approximation used.
Consider a one dimensional wavepacket.
\begin{eqnarray}
u(x,\tau)=\frac{1}{\sqrt{2\pi}}\int_{-\infty}^{\infty} a(k) e^{i(kx-\omega \tau)}dk \label{eee1}
\end{eqnarray}
Here $a(k)$ is the weight of the $k$th component $e^{i(kx-\omega \tau)}$.
$ a(k) $ can be determined as
\begin{eqnarray}
a(k)=\frac{1}{\sqrt{2\pi}}\int_{-\infty}^{\infty} u(x,0) e^{ikx}dx \label{eee2}
\end{eqnarray}
Suppose
\begin{eqnarray}
u(x,0)=e^{-\sigma^2 x^2 +ik_0 x} \label{eee3}
\end{eqnarray}
Then
\begin{eqnarray}
a(k)=\frac{1}{\sigma \sqrt{2}}e^{\frac{(k-k_0)^2}{4\sigma^2}} \label{eee4}
\end{eqnarray}
Therefore, substituting in Eq. \ref{eee1}
\begin{eqnarray}
u(x,\tau)=\frac{1}{\sqrt{2\pi}}\int_{-\infty}^{\infty} dk \frac{1}{\sigma \sqrt{2}} e^{\frac{(k-k_0)^2}{4\sigma^2}} e^{i(kx-\omega \tau)} \label{eee5}
\end{eqnarray}
Therefore,
\begin{eqnarray}
u_{tr}(x+L,\tau+\tau_0 +\Delta \tau)=\frac{1}{\sqrt{2\pi}}\int_{-\infty}^{\infty} dk \frac{1}{\sigma \sqrt{2}} t(k) e^{\frac{(k-k_0)^2}{4\sigma^2}} 
e^{ik(x+L)-i\omega(\tau+\tau_0 +\Delta \tau)} \label{eee6}
\end{eqnarray}
Here $ t(k) $ is transmission amplitude of the length $ L $. $ \tau_0 $ is the time taken to transmit in the absence of potential in the region of length $L$, 
i.e. when $ t(k)=e^{ikL} $.
$ \tau_0 +\Delta \tau $ is the time taken to transmit in the presence of potential.
Therefore,
\begin{eqnarray}
u_{tr}(x+L,\tau+\tau_0 +\Delta \tau)=\frac{1}{\sqrt{2\pi}}\int_{-\infty}^{\infty} dk \frac{1}{\sigma \sqrt{2}} \vert t(k)\vert e^{i\eta_{k}} e^{\frac{(k-k_0)^2}{4\sigma^2}} 
e^{ik(x+L)-i\omega(\tau+\tau_0 +\Delta \tau)} \label{eee7}
\end{eqnarray}
where $t(k)= \vert t(k) \vert e^{i\eta_{k}}$.
Now if $ \sigma^{2} $ is very small then $ \vert t(k)\vert $ can be taken to be independent of $ k $ and $ \vert t(k)\vert =t $.
Thus if dispersion of wavepacket becomes stronger in the low energy
quantum regime we have to take fewer and fewer momentum states to make the wavepacket and this may not be possible
as we go upto the low energies where $\Delta \tau$ may become negative in 1D. So a clear meaning of negative WDT has not emerged till date. From \ref{eee7} 
\begin{eqnarray}
u_{tr}(x+L,\tau+\tau_0 +\Delta \tau)=\frac{1}{2\sqrt{\pi}}\frac{1}{\sigma}t\int_{-\infty}^{\infty}dk e^{\frac{(k-k_0)^2}{4\sigma^2}} 
e^{i\eta_{k}+ik(x+L)-i\omega(\tau+\tau_0 +\Delta \tau)} \label{eee8}
\end{eqnarray}
So for the wavepacket to remain undispersed the weight of a particular component must remain unchanged implying,
\begin{eqnarray}
kx-\omega \tau=\eta_{k}+k(x+L)-\omega(\tau+\tau_0 +\Delta \tau)=\text{constant} \label{eee9}\\
\text{or,}\hspace*{.2cm} \eta_{k}+kL-\omega(\tau_0 +\Delta \tau)=0 \nonumber\\
\text{or,}\hspace*{.2cm} \frac{d\eta_{k}}{d\omega}+L\frac{dk}{d\omega}=\tau_0 +\Delta \tau \label{eee10}
\end{eqnarray}
Also in the absence of scatterer
\begin{eqnarray}
\therefore \hspace*{.2cm} kx-\omega \tau&=&k(x+L)-\omega(\tau+\tau_0)\nonumber\\
\text{or,}\hspace*{.2cm} kL -\omega \tau_0&=&0\nonumber\\
\text{or,}\hspace*{.2cm} L\frac{dk}{d\omega}&=&\tau_0 \label{eee11}
\end{eqnarray}
Therefore from Eq. \ref{eee10}
\begin{eqnarray}
\frac{d\eta}{d\omega}&=&\Delta \tau\nonumber\\
\frac{d\eta}{dE}\frac{dE}{d\omega}&=&\Delta \tau\nonumber\\
\hbar \frac{d\eta}{dE}&=&\Delta \tau^W \hspace*{2cm}\text{as}\hspace*{.2cm} E=\hbar\omega \label{eee12}
\end{eqnarray}

\begin{figure}[h]
\centering
\includegraphics[scale=0.35]{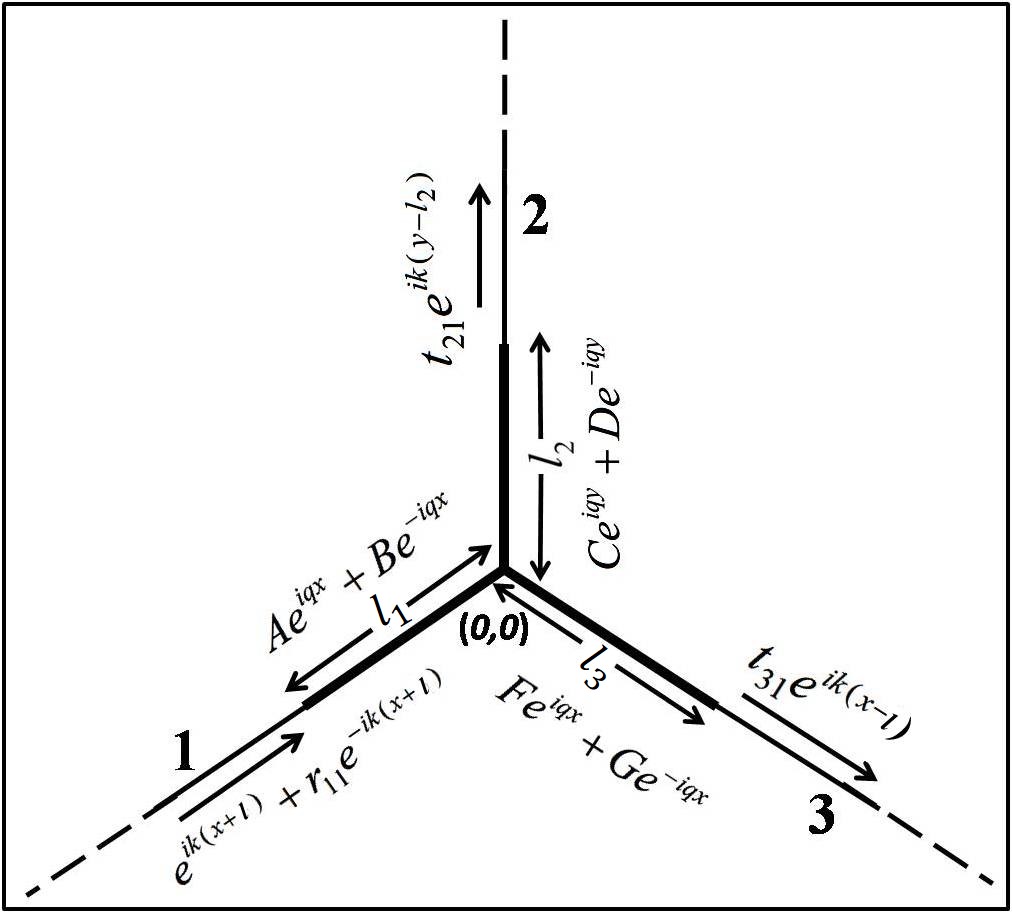}
\caption{\label{fig1}Schematic representation of scattering of electrons by a three prong potential. The three arms labelled as $ 1,2,3 $ are one dimensional quantum 
wires meeting at point marked $ (0,0) $. The thick solid lines represents quantum wires with finite potential $ V $. The thin solid lines represents quantum wires 
without any potential, i.e. $ V=0 $. The electron propagation direction is shown by arrows. The
dashed lines represent the fact that the quantum wires are connected to 
electron reservoirs via leads. $ r_{11} $ is the amplitude of electrons incident from $ 1 $ and reflected back to $ 1 $. $ t_{21} $ is the transmission amplitude for 
electrons incident from $1$ and transmitted to $2$ and $ t_{31} $ is the transmission amplitude for electrons incident from $ 1 $ and transmitted to $ 3 $.
The wavefunction in the different regions and the lengths of the three legs $l_1$, $l_2$ and $l_3$ are shown in the figure.}
\end{figure}

All wavepackets in 1D, 2D and 3D, disperse and so Eq. \ref{eee9} 
correspond to an approximation called stationary phase approximation. Hence in a quantum regime if stationary
phase approximation fails then Eq. \ref{eee12} will not give
the traversal time while at energies larger than the scale at which the potential acts (de-Broglie wavelength is larger than the spatial scale in which
the potential varies) it correctly gives the traversal time. 
The treatment can be naturally extended to higher dimensional S-matrices. Consider the three prong potential \cite{urba}
shown in Fig. \ref{fig1} and explained in details in the figure
caption.
$ k=\sqrt{\frac{2mE}{\hbar^2}} $ and $ q=\sqrt{\frac{2m(E-V)}{\hbar^2}} $ are wave vectors along the thin lines and the thick solid lines respectively. $ r_{11} $ 
is the amplitude of electrons incident from $ 1 $ and reflected back to $ 1 $. $ t_{21} $ is the transmission amplitude for electrons incident from $1$ and transmitted 
to $2$ and $ t_{31} $ is the transmission amplitude for electrons incident from $ 1 $ and transmitted to $ 3 $. An incident wavepacket along lead $1$ will split into
three wavepackets that are reflected and transmitted to leads $2$ and $3$. The WDT $\Delta \tau_{31}^W$ for the wavepacket coming out of lead 3 will be 
\begin{equation}
\hbar \frac{d\theta_{t_{31}}}{dE}=\Delta \tau_{31}^W \label{eee14}
\end{equation}
Here $\theta_{t_{31}}$ is the scattering phase shift of the electrons going from $1$ to $3$.
Note that the propagation of a wavepacket in time has been used to derive Eq. \ref{eee14}. 
An axiom in quantum mechanics states that if we project one particle or a small number of particles (typically represented by a wavepacket)
into a scatterer then this particle may get reflected
or transmitted to 2 or to 3 in a random manner. But 
if we repeat the process with a statistically large number of electrons then it will produce the interference pattern and all other observable quantities 
like $r_{11}$, $t_{21}$ and $t_{31}$ that are obtained 
from the stationary state solutions of the time independent Schrodinger equation. 
In other words, $\eta_k$ in Eq. \ref{eee8} or $\theta_{t_{31}}$ in Eq. \ref{eee14} can be obtained from the time independent Sc. equation.

One can start from the time independent Sc. eqn and derive \cite{butt1, butt2} a traversal time called LPT as
\begin{equation}
\Delta t_{31}^{L}=\hbar \int_{sample} d\vec{r} \Delta \epsilon \frac{\delta\theta_{t_{31}}}{\delta V(\vec{r})} \label{eee15}
\end{equation}
for particles transmitted from 1 to 3.
$|t_{31}|^2$ fraction of the incident electrons get transmitted like this.
Here $\frac{\delta}{\delta V(\vec{r})}$ means a functional derivative with respect to the local potential $V(\vec{r})$.
$\int_{sample}d \vec{r}$ means an 
integration over all the coordinates of the sample or the scattering region where the wavefunction deviates from the typical asymptotic wave functions
of a scttering problem. In
case of Fig \ref{fig1} the sample or the scattering region is simply the region shown in thick lines.
This derivation unlike the derivation of WDT in Eq. \ref{eee12} does not use any approximation. But when WDT is correct, it has to give the same physical
quantity as LPT. So if a signal can be sent in WDT (the correct one) it can be also sent in LPT.

At this point we would like to clearly define the three terms we use in this paper. First is the "traversal time" which will be used in a literary sense that
it is a time associated with a particle going from one point to another point traversing the intermediate states connecting the two points. So for a time interval
to qualify as traversal time
it has to be related to DOS. Second is the "LPT" which is exact and always sums up to give the correct DOS. Third is the "WDT" which is correct if stationary phase
approximation is correct and not otherwise. However, from now on 
we will not use terms like "correct WDT" and "incorrect WDT". Correctness or incorrectness will be born by
its comparison with LPT. If WDT is correct then the Copenhagen interpretation requires it to be equal to LPT.

We will show that in 1D LPT remain positive while WDT becomes negative at low energies.
We use Burgers-circuit (B-C) \cite{berr1, berr2}
to prove this. Next we will show that in Q1D, one can find frequently occurring regimes where LPT and WDT can both be
identically negative. That is we will prove that in Q1D there are regimes wherein
\begin{equation}
\Delta t_{31}^L=\Delta t_{31}^W < 0 \label{eee16}
\end{equation}
Before showing this we would like to summarize why such a demonstration is interesting. First of all it is new and no such previous examples or a system that exhibit
this can be cited. Secondly it implies that $\Delta t_{31}^L$ can be quantitatively negative and such a quantitative value for it has not been obtained before.
$\Delta t_{31}^L < 0$ does not prove anything with respect to signal propagation time
unless one can show $\Delta t_{31}^L=\Delta {t_{31}}^W < 0$ because only then one can claim
that a wavepacket or some change in spatial probability distribution of electrons can be transmitted to negative times within the single particle
coherence length.

Since quantum mechanics and relativity are anyway not consistent with each other but both
very successful in their own regimes it has been always believed that superluminal times are possible in quantum mechanics. But a
concrete demonstration consistent with quantum mechanics (no need to bring relativity in the picture) itself has not been shown so far. 
We will use the three prong potential to illustrate our results.
The proof is general, and can be applied to any scattering matrix element $S_{\alpha \beta}$ that makes loops in the first Riemann
surface and in mesoscopic systems one can find a large class of systems that exhibit this.
Burger's circuit uses the analyticity of the complex scattering matrix elements and is therefore more general than
quantum mechanics and so quantum mechanics has to respect it.
Argand-diagram (A-D) is a plot of imaginary versus real parts of an analytic complex function like transmission amplitude $t(k)$ 
and Burgers circuit states that for a closed contour $C$ in the A-D \cite{berr1, berr2}.
\begin{equation}
\oint_{C} d\phi=2\pi I
\label{eee17}
\end{equation}
where, $ \phi=Arctan\frac{Im(t)}{Re(t)} $. At the origin $Re(t)$=$Im(t)$=0 implying the origin is a singular point in the complex plane.
If the contour $ C $ does not enclose this phase singularity then $I$ is 0. When the contour $ C $ enclosing a phase singularity 
is clockwise then $I$ is -1 and when the contour $ C $ enclosing a phase singularity is counter-clockwise then $I$ is +1.

\begin{figure}[h]
\centering
\includegraphics[scale=0.35]{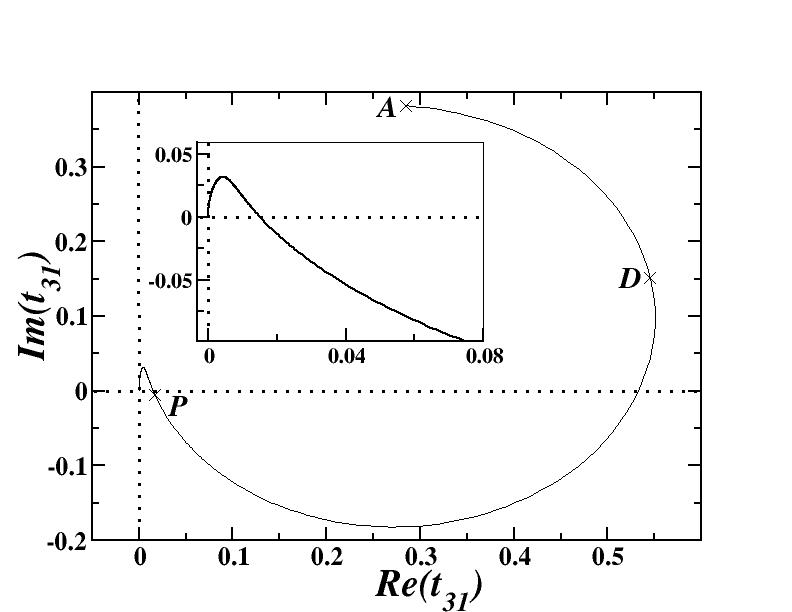}
\caption{\label{fig2}Argand diagram for transmission amplitude $ t_{31} $ for the three prong potential shown in Fig. \ref{fig1}. The Argand diagram is obtained by 
varying the wavevector $ kl $ from $0$ to $5$. Here, $ l_{1}=l_{3}=l, l_{2}=5l $ and $ eVl=-1000 $. The inset show a magnified picture near the origin.}
\end{figure}

\begin{figure}[h]
\centering
\includegraphics[scale=0.35]{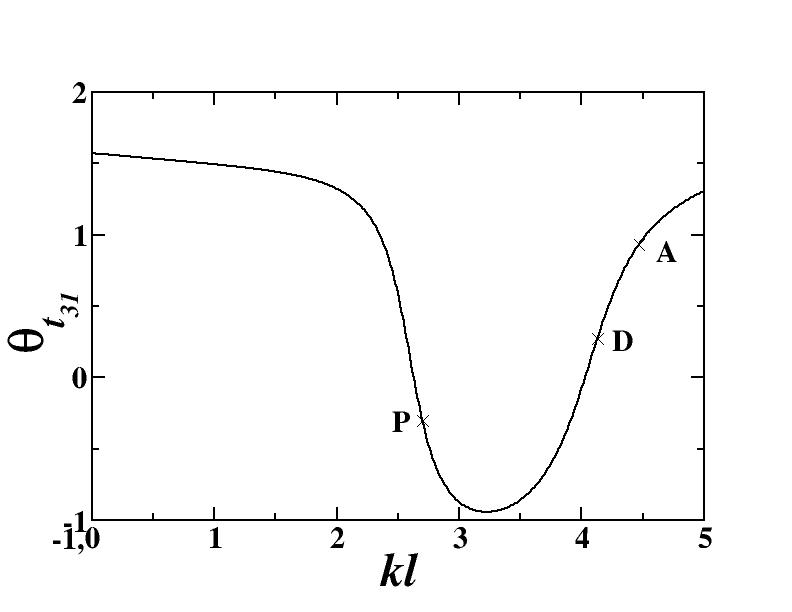}
\caption{\label{fig3}Plot of transmission phase shift $ \theta_{t_{31}} $ corresponding to transmission amplitude
$t_{31}$ for the three prong potential shown in Fig.\ref{fig1}, as a function of $ kl $. Here, $ l_{1}=l_{3}=l, l_{2}=5l $ and $ eVl=-1000 $.}
\end{figure}

\begin{figure}[h]
\centering
\includegraphics[scale=0.35]{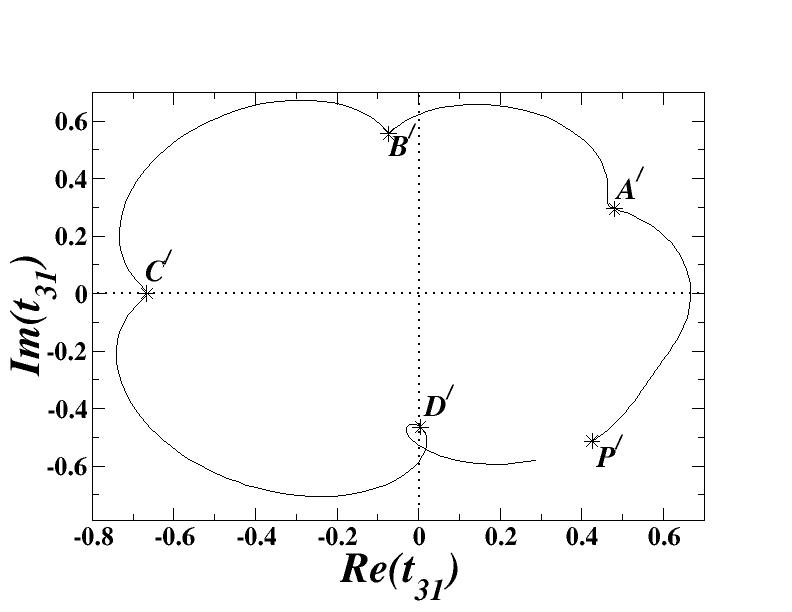}
\caption{\label{fig4}Argand diagram for transmission amplitude $ t_{31} $ for the three prong potential shown in Fig. \ref{fig1}.
The Argand diagram is obtained by varying the potential $ eVl $ from $0.0$ to $-25$. 
It starts from the point marked P'. Here, $ l_{1}=l_{3}=l, l_{2}=5l $ and $ kl=2.7 $ which are the parameters
corresponding to the point P in Figs. 2 and 3. Even when $kl$ value is close to zero rather than 2.7, the point P' does not get close to the
origin. The phase singularity is at origin at which $ t_{31}=0 $.}
\end{figure}

\begin{figure}[h]
\centering
\includegraphics[scale=0.35]{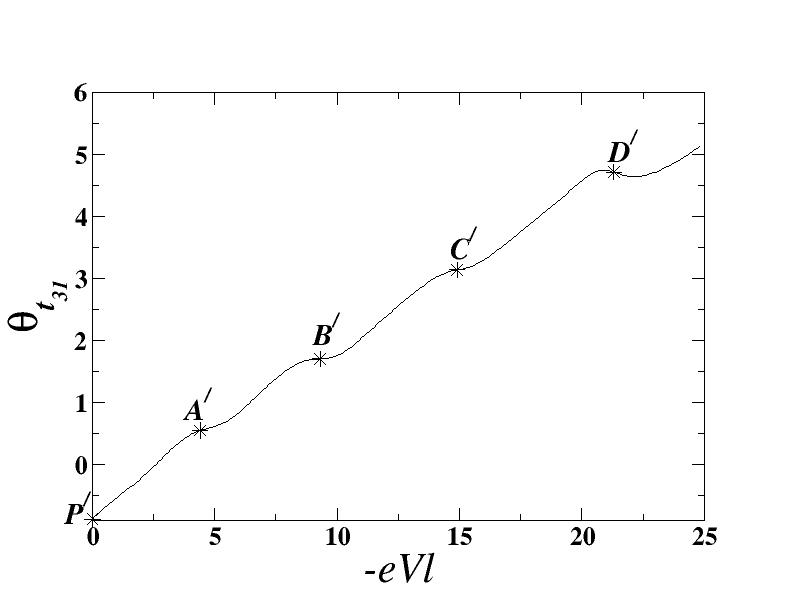}
\caption{\label{fig5}Plot of transmission phase shift $ \theta_{t_{31}} $ corresponding to the A-D in Fig. 4,
as a function of $-eVl $ which is positive as $V$ is negative.}
\end{figure}

In Fig. \ref{fig2} 
we draw the A-D for $t_{31}$ and in Fig. \ref{fig3} 
we plot $\theta_{t_{31}}$ as a function of the wavevector for the same choice of parameters. 
Note that in Fig. \ref{fig3} around the point marked P, the scattering
phase shift decreases with $kl$ that is $d \theta_{t_{31}} \over dE$ will be negative. 
The corresponding point in the A-D is also
marked $P$ in Fig \ref{fig2}. One can calculate $\theta_{t_{31}}=
Arctan[\frac{Im (t_{31})}{Re (t_{31})}]$ at point $P$ in Fig. \ref{fig2} and close by points to see from the Argand diagram as well that
$\theta_{t_{31}}$ has a negative slope here. This negative slope does not have anything to do with the
three prong potential and it can be observed for all values of $l_2$ at the same energy and of the same magnitude. For example by making $l_2 \rightarrow 0$ 
in Fig. 1, we will
get the 1D limit where too this negative slope can be seen at the same energy and of the same magnitude. This negative slope in 1D has been the matter of 
study in the past and there is no conclusive understanding as to whether it means something physical or is an artefact of the stationary phase
approximation \cite{land}. 
It can be seen in a one dimensional potential like a square barrier or a square well potential where it happens only once as the energy or the momentum
is increased from 0.
One can understand the origin of this negative slope from the A-D using B-C. At zero energy the transmission is zero even for an infinitesimal potential in the
scattering region, implying
$Im (t_{31})$ and $Re (t_{31})$ are both zero. This is a point where $\theta_{t_{31}}$ is singular. So the A-D starts or originates from the singularity and comes out
of the singularity. Now since the absolute value of $t_{31}$ is bounded the A-D has to curl around the singularity either in clockwise direction or anticlockwise direction.
Clockwise will mean phase will decrease whereas anticlockwise will mean phase will increase with energy as suggested by Eq. \ref{eee17}.
Initially however, when the A-D trajectory emerges from the singularity,
it is neither clockwise nor anticlockwise with respect to it. After a while it turns around and before becoming anticlockwise with respect to the origin it goes
through a small region around point P where it is clockwise, and this is what creates the negative slope.
Once it has gone anticlockwise it never turns back again in 1D. 
And so after this initial drop phase monotonously increases. The phase of the wavefunction
increases implies rotation of the wavefunction in Hilbert space unidirectionally which is like a notion of time increasing monotonously and positively, as
time evolution is guided by a factor $e^{iEt/\hbar}$.
Now in case of the three prong potential we would like to make the following observations that for some reason has not been observed in 1D essentially because
making these observations is not very revealing in 1D. Our observations for the three prong
potential will become more relevant as the paper progresses. First of all one can keep the energy (or $kl$)
fixed at the value corresponding to the point $P$ in
Fig \ref{fig2}, and decrease (or increase)
the potential globally by constant amounts to generate identically same A-D starting from P. Energy increasing corresponds to potential decreasing globally and vice
versa. But this is a very obvious observation. Now to come to the point suppose we keep the energy fixed at that corresponding to point $P$ and decrease the potential
only in the thick region in Fig. \ref{fig1} (that is only in the scatterer) by constant amounts starting from 0. Unlike the A-D
of Fig. \ref{fig2} this A-D will not start from the singularity but start from a point completely different from P. 
Because when the potential is zero, even a mode with infinitesimal energy or momentum gets fully transmitted.
So even if the point P in Fig. 2, The A-D obtained by varying the potential $V$ will not start from the origin.
This A-D is shown in Fig \ref{fig4}. It starts from $\text{P}^/$ and right from the very start it goes anticlockwise with respect to the origin
or with respect to the singularity, its trajectory being completely unaffected by the presence of the singularity.
The scattering phase shift shown in Fig \ref{fig5} corresponding to this A-D, therefore starts increasing from 0 and there is no negative slope near the origin.
Only at a higher value of the
potential $V$ i.e., at the point $\text{D}^/$ it shows a negative slope
and we will come back to this later.
Decreasing the constant
potential $V$ only in the thick region in small steps of $e\Delta \epsilon$ allows us to evaluate the RHS of Eq. \ref{eee15} and thus $\Delta t_{31}^L$.
Here $e$ is the particle charge that is set to 1 without any loss of generality.
Thus this A-D explains why LPT will not be negative at P.
So at low energies or in the quantum regime, global constant shift in potential and constant shift of potential in the scattering region
produce completely different results. 
Adding up all possible LPT for all the channels like $t_{21}$, $r_{11}$ etc., 
and multiplying with $\frac \pi \hbar$ one can get density of states (DOS) that can be determined from
the internal wavefunction also for comparison. We have checked that there is perfect agreement of this LPT with the DOS determined from internal wavefunction.
In Fig. 6 we plot $\frac{d \theta_{t_{31}}(E)}{dE}$ and $\frac{d \theta_{t_{31}}(E)}{deV}$ 
where $V$ is the constant potential in the sample region. Both are plotted as a function of $kl$. It shows 
$\frac{d \theta_{t_{31}}}{dE}$ is strongly negative near the origin but $\frac{d \theta_{t_{31}}}{deV}$ is not.
This negativity of $\frac{d \theta_{t_{31}}}{dE}$ is due to the failure of stationary phase approximation as it does not agree with LPT. 
LPT starts from zero and starts going negative from the very start (note that both the derivatives are calculated at a large value for
potential $V$ in the sample) and this is a speciality of Q1D that we will explain soon. However, since it does not agree with WDT Eq. \ref{eee16}
is not satisfied. However, at three broad
minima shown in Fig. 6, Eq. \ref{eee16} is satisfied. There are other regions also where $\frac{d \theta_{t_{31}}}{dE}$ and 
$\frac{d \theta_{t_{31}}}{deV}$ are identical but since they are not negative there Eq. \ref{eee16} is not satisfied. At still higher values of $kl$
we do not find negative regimes.
We will come back to this later.

Now in case of Fig \ref{fig2}, 
if we increase the energy further then the A-D show subloops within a particular Riemann surface and demonstrated
in Fig. \ref{fig7}. At each subloop we get a portion of A-D trajectory that is clockwise with respect to the origin. For example we have highlighted
one such subloop in Fig. \ref{fig7} by dotted lines and marked the subloop as MNOM. Between N and O the trajectory is clockwise or in other words
the normal vector to the curve points towards to origin and scans the origin. Between N and O $\frac{d \theta_{t_{31}}}{dE}$ will be negative.
The identical A-D of Fig. 7
can be generated by keeping the energy fixed at the value corresponding to point P and decreasing (or increasing) the potential globally by constant amounts. 
We have already discussed 
that in a scattering problem increasing incident energy by $ dE $ is equivalent to decreasing the potential globally by a constant amount $ \Delta\varepsilon $,
such that $ dE=-e\Delta\varepsilon $, where $ e $ is particle charge that we will set to $ 1 $ to simplify our arguments. That is, the changed potential is
$ V'(\vec{r})=V(\vec{r})-\Delta\varepsilon $. Hence if we can generate a closed sub-loop in the Argand diagram by varying $E$
in small steps of $dE$, then we can also do so by globally changing the potential in small steps of $\Delta \epsilon$. Decreasing
the potential globally is equivalent to increasing the energy and so it will generate the A-D further away from the origin and increasing the potential is equivalent
to decreasing the energy and so it will generate the A-D from P upto the origin. So the subloops can also be generated by decreasing the potential globally 
starting from the $V$ value used in Fig. 7 and keeping the energy fixed at the value corresponding to point P in Fig. 7.
(globally means in the thick regions as well as in the leads in Fig. \ref{fig1}). 
One particular subloop MNOM is depicted in Fig. \ref{fig7}.
For a closed contour A corresponding to a particular subloop in Fig. \ref{fig7},
\begin{eqnarray}
\oint_{A} \Delta\theta_{s_{\alpha\beta}}=0 \nonumber\\
\oint_{A} \frac{\Delta\theta_{s_{\alpha\beta}}}{\Delta E}dE=0 \nonumber\\
\text{i.e.}\hspace{1cm} \oint_{A} \frac{\Delta\theta_{s_{\alpha\beta}}}{\Delta E}dE= 
-\oint_{A}\int_{global}\frac{\Delta\theta_{s_{\alpha\beta}}}{\delta V(\vec{r})}\Delta\varepsilon d\vec{r}=0\label{eee18}
\end{eqnarray}
Here $\int_{global} d\vec{r}$ correspond to an integration over all spatial coordinates.
These subloops are consistent with B-C and hence the topological structure of the Riemann surface in which the
A-D is drawn. Subloops also belong to a different topological class such that a particular subloop is very fundamental and cannot be removed by varying any parameter. 
Essentially it means a multiply connected curve cannot be continuously deformed to a simply connected curve. Physically, these subloops originate from
Fano resonance and a resonant state cannot be removed by varying any parameter. A resonant state can be populated or de-populated but the total number of states in a
system remain conserved. So if we do not increase the incident energy (or decrease the potential globally) but decrease the potential only in the thick region of
Fig. \ref{fig1} then we will generate identical number of subloops. The nature of the different subloops may change but the total number of subloops will be the same. This is
shown in Fig. \ref{fig8} implying that subloops are very common and will persist upto very high energies (such high energies are not shown in Fig. \ref{fig7}).
Therefore, for a closed contour A one can find the corresponding closed contour B in Fig. \ref{fig8}, such that,
\begin{eqnarray}
-\oint_{B}\int_{sample}\frac{\Delta\theta_{s_{\alpha\beta}}}{\delta V(\vec{r})}\Delta\varepsilon d\vec{r}=0 \label{eee19}
\end{eqnarray}
By the argument that if a subloop A can be generated by varying the incident energy $E$ then a corresponding subloop $B$ can be generated by varying the potential
uniformly in the sample region.

This only proves that the integrations are equal, i.e.,. 
\begin{equation}
\oint_{A} \frac{\Delta\theta_{s_{\alpha\beta}}}{\Delta E}dE=
-\oint_{B} \int_{sample}\frac{\Delta\theta_{s_{\alpha\beta}}}{\delta V(\vec{r})}\Delta\varepsilon d\vec{r}=0 \label{eee20}
\end{equation}
This also proves that the integrands are negative in part of the range in which the relevant parameter is varied to get the closed contours A and B
and positive in the rest of it. Thus this proves that LPT which is always correct, can be negative.
All the subloops A in Fig. \ref{fig7} and subloops B in Fig. \ref{fig8} satisfy Eq. \ref{eee20}.
But this does not explain that around three minima in Fig. 6 at negative values why Eq. \ref{eee16} should be valid.
To explain that we have to show that the integrands can be equal in some regimes. If we can find some regime
where the value of the integration becomes independent of the shape and size of the contour A as well as B, then the integrands have to be identical.
And also unlike the subloops in Fig. \ref{fig7} and Fig. \ref{fig8} the trajectory making subloops A and B
should not have any discontinuities but should come back onto itself very smoothly or else the integration limits (like the point M in Fig. 7
in the trajectory M to N is not identical to the point M in the trajectory O to M) are not identical in all respect.
If closed contour A satisfy
this property then closed contour B too has to satisfy this property. Because the LHS integration cannot be 0 independent of the shape and size of the contour A
while the RHS integration does depend on the shape and size of the contour B.
Such smooth subloops A are shown in Fig. \ref{fig9} and smooth subloops B are shown in Fig. 10. 
Relevant parameters are mentioned in the figure caption. Note that in Fig. \ref{fig10} we have fixed the energy 
such that $kl$=8.22 and the initial value of $eVl$=-1000 which implies the trajectory will start exactly from the point marked x in Fig. \ref{fig10}, which is
identical to the point at $kl$=8.22 in Fig. \ref{fig9}.

In Fig.11 we plot $\frac{d \theta_{t_{31}}(E)}{dE}$ and $\frac{d \theta_{t_{31}}(E)}{deV}$ as a function of $kl$ and above a value of $kl$=8.22 the two curves become
identical as expected from Figs. 9 and 10. There are broad energy ranges occurring periodically where Eq. \ref{eee16} is satisfied and this happens upto
very high energies. Obviously, if we construct
a wavepacket with the modes in these ranges then that wavepacket will remain undispersed perfectly satisfying the stationary phase approximation. This is because
the A-D in these energy ranges make periodic orbits and all the relevant quantities in the integrand in Eq. \ref{eee8} vary in a periodic fashion, the positive
parts balancing the negative parts. In case of Fig. 6, by the time we reach $kl$=7.5 the dispersion due to scattering by the $l_1$ and $l_3$ legs 
of Fig. 1, become negligible
and we get perfect agreement between $\frac{d \theta_{t_{31}}(E)}{dE}$ and $\frac{d \theta_{t_{31}}(E)}{deV}$. Although they become mostly positive, they do
show three very broad energy ranges where Eq. \ref{eee16} is satisfied.

In Fig. 12 we plot $\theta_{t_{31}}$ as a function of $kl$ corresponding to the A-D of Fig. 9. Note the periodically occurring regimes where 
$\frac{d \theta_{t_{31}}(E)}{dE}$ is negative. For example we can
take the range $kl$= 8.5 to 8.9 and construct a wavepacket with these modes by choosing
$a(k)=\frac{1}{\sigma \sqrt{2}}e^{\frac{(k-k_0)^2}{4\sigma^2}}$ sharply peaked at $k_0=$8.7 and decaying beyond 8.5 and 8.9 by a suitable choice of
$\sigma$. We can create this wavepacket arbitrarily close or at $x=0$ in Fig. 1 as $l_1=l_3$=0 in Fig. 12. At the next instant of time this wavepacket will be displaced
by a WDT along the time axis which is negative. So the electrons in the wavepacket will move in negative time and so they will behave like positrons.
A positron will attract an electron and annihilate each other to form a new particle or quasi particle. High energy phenomenon of producing photons
will not occur in quantum wires as it happens for real positrons. However, this new quasi particle will be quite stable due to the fact that
negative times are coming from the solution of the fundamental equations of motion, and so they will move as a bound state.

In conclusion, we wanted to demonstrate Eq. \ref{eee16} for certain range of energies and we have successfully shown that in Fig. 6 as well as in Fig. 11.
This also happens for another potential that can be solved exactly in Q1D, that of a negative delta function potential in a multichannel quantum wire.
We have checked this for a two channel quantum wire. Physical consequence of Eq. \ref{eee16} is that a signal can be propagated in negative time. We also 
conjecture that one can have a bound state of two electrons as a consequence.

\begin{figure}[h]
\centering
\includegraphics[scale=0.35]{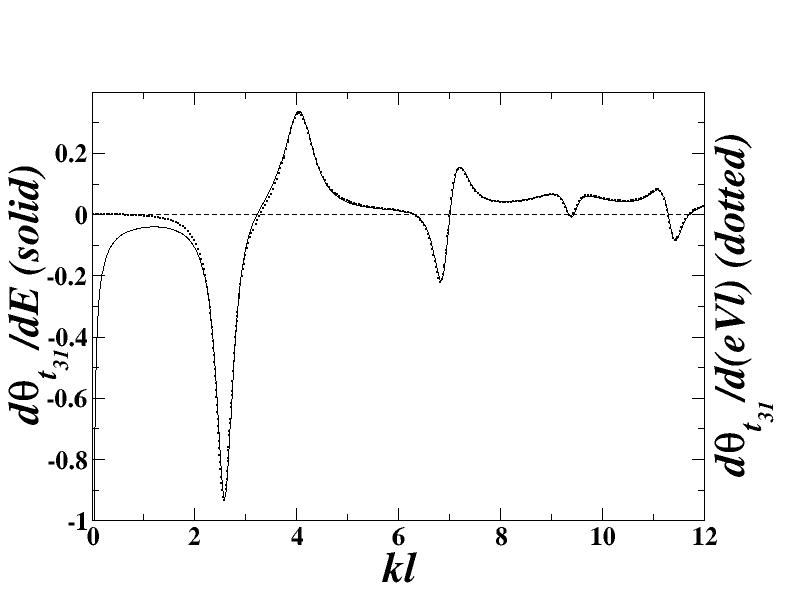}
\caption{\label{fig6} Here, $ l_{1}=l_{3}=l, l_{2}=5l $ and $ eVl=-1000 $.}
\end{figure}
\begin{figure}[h]
\centering
\includegraphics[scale=0.35]{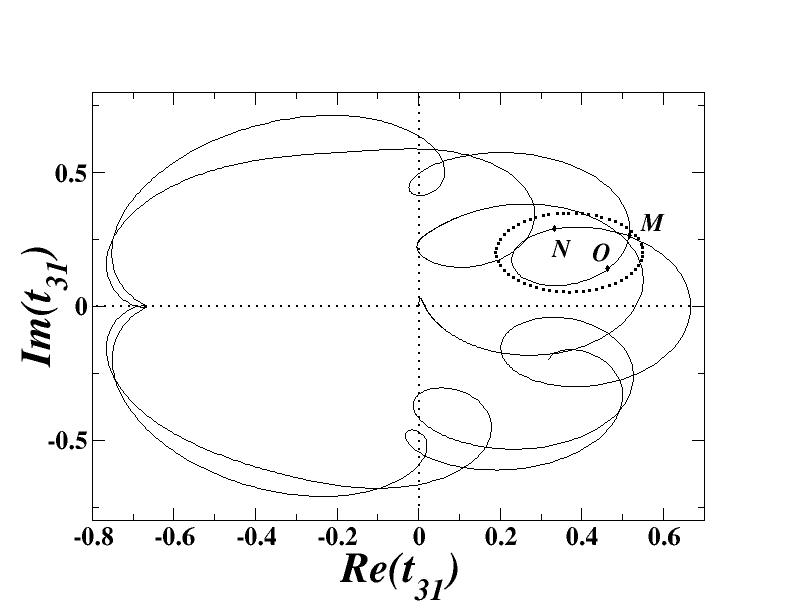}
\caption{\label{fig7} $ kl $ is varied from 0 to 20. to get this A-D. Here, $ l_{1}=l_{3}=l, l_{2}=5l $ and $ eVl=-1000 $.}
\end{figure}
\begin{figure}[h]
\centering
\includegraphics[scale=0.35]{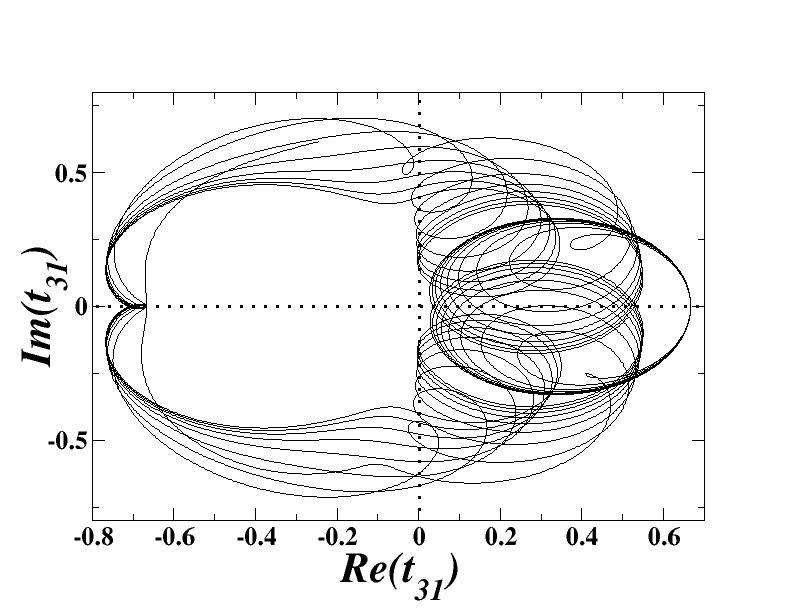}
\caption{\label{fig8}Argand diagram for transmission amplitude $ t_{31} $ for the three prong potential shown in Fig. \ref{fig1}. The Argand diagram is obtained by 
varying the potential $ eVl $ from $-1$ to $-1000$. Here, $ l_{1}=l_{3}=l, l_{2}=5l $ and $ kl=4 $.}
\end{figure}

\begin{figure}[h]
\centering
\includegraphics[scale=0.35]{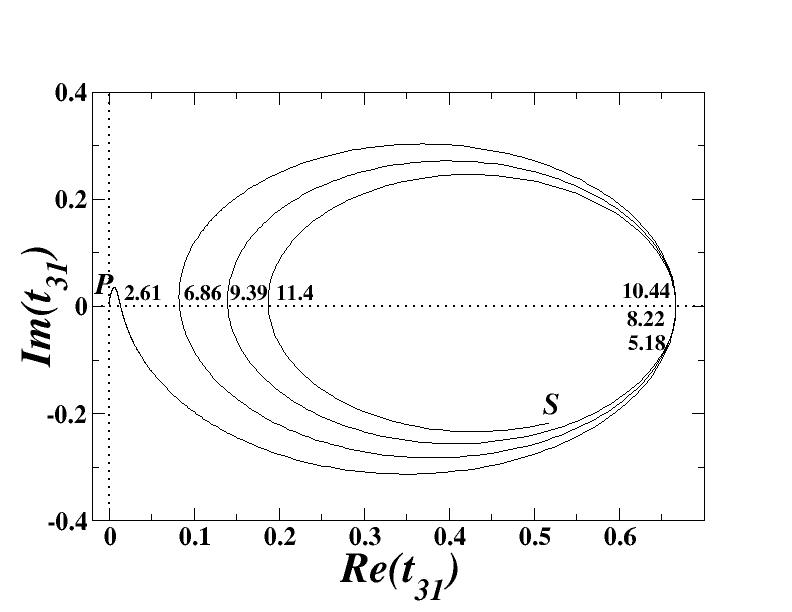}
\caption{\label{fig9}A-D for $t_{31}$ for the system in Fig. 1. Here $l_1=l_3=0$, $l_2$=5, and $eVl$=-1000. $kl$ is varied from 0 to 12.5 and some $kl$ values
like 2.61, 5.18, 6.86, etc are marked at corresponding points in the A-D. It means the first time the A-D crosses the real axis is at $kl=2.61$. Then it crosses
again at 5.18, then at 6.86 and so on.}
\end{figure}
\begin{figure}[h]
\centering
\includegraphics[scale=0.35]{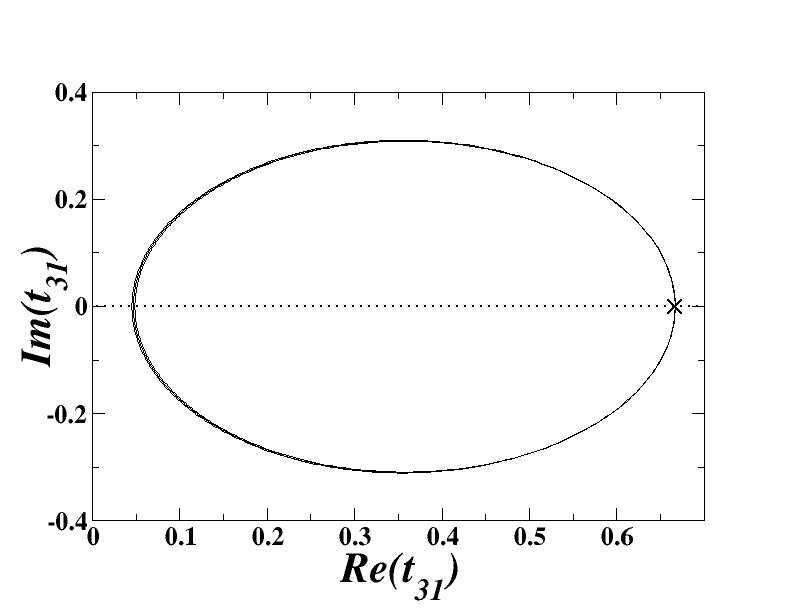}
\caption{\label{fig10}A-D for $t_{31}$ for the system in Fig.1. $l_1=l_3=0$, $l_2=5$, and $kl$= 8.22. $eVl$ is varied from -1000 to -1050. Note that the
starting point is marked with a x which is the same are the point marked 8.22 in Fig.9 as the parameter values are the same.}
\end{figure}
\begin{figure}[h]
\centering
\includegraphics[scale=0.35]{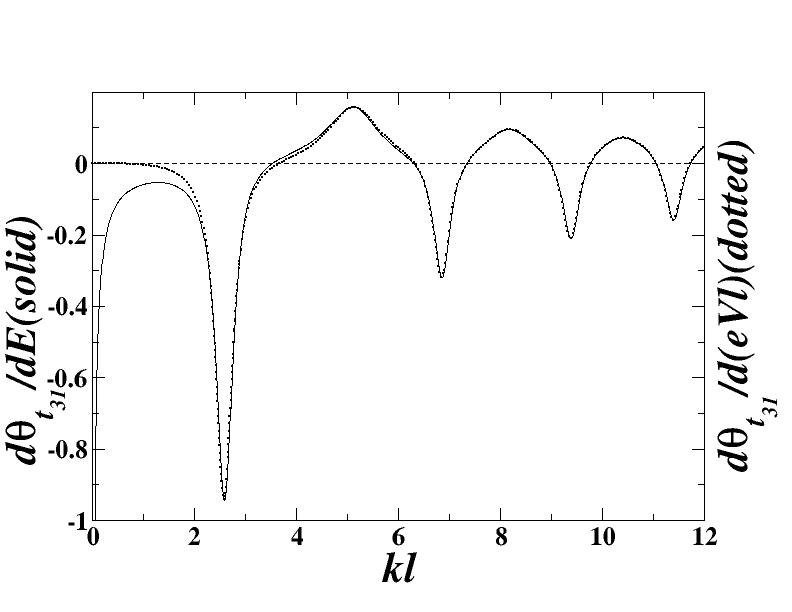}
\caption{\label{fig11}Here $l_1=l_3=0$, $l_2=5$ and $eVl$=-1000.}
\end{figure}
\begin{figure}[h]
\centering
\includegraphics[scale=0.35]{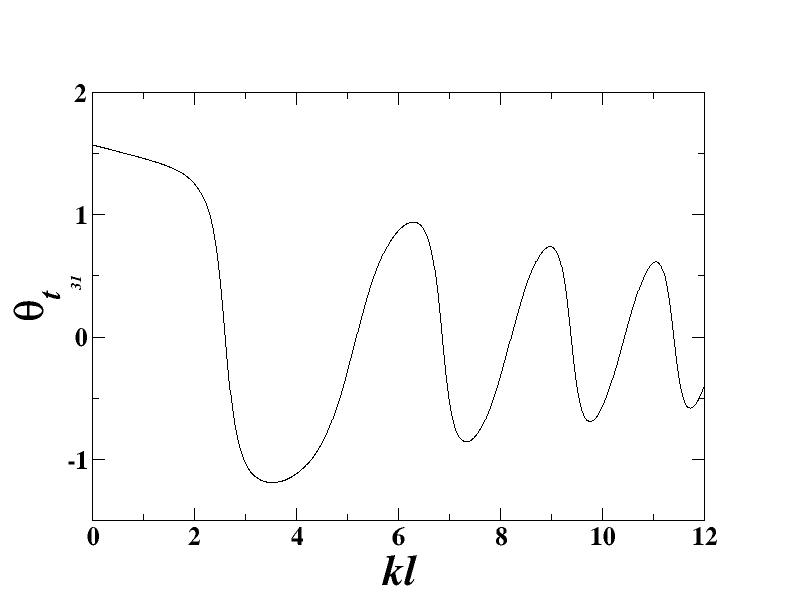}
\caption{\label{fig12}Here $l_1=l_3=0$, $l_2$=5, $eVl$=-1000. }
\end{figure}

\end{document}